\documentclass[runningheads,a4paper]{llncs}

\usepackage{amssymb}
\usepackage{graphicx}

\usepackage{url}

\usepackage{amsmath}
\usepackage{tikz}
\usetikzlibrary{calc}

\usepackage{caption,subcaption}
\usepackage{multirow}

\usepackage{booktabs}

\usepackage{epstopdf}

\usepackage{listings}

\begin{document}

\mainmatter  

\title{Even better correction\\ of genome sequencing data}

\titlerunning{Even better correction of genome sequencing data}

%
%
\author{Maciej D{\l}ugosz\and Sebastian Deorowicz\and Marek Kokot}
\authorrunning{Maciej D{\l}ugosz\and Sebastian Deorowicz\and Marek Kokot} %

\institute{Institute of Informatics, Silesian University of Technology,\\
Akademicka 16, 44-100 Gliwice, Poland\\
\path|{maciej.dlugosz,sebastian.deorowicz,marek.kokot}@polsl.pl|\\
\url{http://sun.aei.polsl.pl/REFRESH/}}

%
%

\maketitle

\begin{abstract}
We introduce an improved version of RECKONER, an error corrector
for Illumina whole genome sequencing data. By modifying
its workflow we reduce the computation time even 10 times. 
We also propose a new method of
determination of $k$-mer length, the key parameter of $k$-spectrum-based family of correctors. 
The correction algorithms are examined on huge data sets, i.e., human and maize genomes for both
 Illumina HiSeq and MiSeq instruments.
\end{abstract}

\section{Introduction}

For several years next-generation sequencing (NGS) technologies, like
454 pyrosequencing, Complete Genomics, Illumina, Ion Torrent, Pacific Biosciences~\cite{LBM16}, and Oxford Nanopore \cite{Nanopore}
have been dominating the field of DNA analysis. 
Their advantages over Sanger method \cite{SNC77}, that they superseded, include extremely high throughput (even $600\,\text{Gb}$ per instrument run), low sequencing cost (even $41\text{\$}$ per Gb) \cite{Q12}
and, in case of Oxford Nanopore MinION, virtually domestically low cost ($1000 \text{\$}$)
and size of the instruments \cite{Nanopore}.

Such advantages have enabled many sequencing data applications.
High availability of huge amount of data have enormous
practical potential.
\emph{De novo} assembly, reassembly, single nucleotide polymorphisms (SNPs)
detection, metagenomics, personalized medicine analysis are only the
examples of applications variety. Large projects aiming of sequencing
thousands of large genomes 
like the 1000 Genomes Project \cite{S08} or the Genome 10K Project~\cite{H09}
became possible.

The price of NGS advantages is relatively low quality of reads (short fragments of sequenced genomes) they produce. 
This poses
a challenge in development of algorithms performing downstream
analysis. Some instruments produce short reads, e.g.,
reads generated by popular Illumina HiSeq2000 of length up to $150\,\text{bp}$ (base pairs) \cite{Q12} are shorter
than many of repeats present in DNA. The distribution of reads over
the genome is far from uniform, which causes overrepresentation of some
fragments and often unpresence of the other ones \cite{LBM16,Q12}. Sequencing errors
cause deformations of reads by altering some symbols, inserting
alien fragments into reads, or removing their fragments \cite{LBM16}.

All of those and other issues have negative impact on utilization of
sequencing data. They affect accuracy of algorithms processing the reads,
e.g., by generating false positives in SNPs detection or causing false
connections between contigs in \emph{de novo} assembly. The algorithms
processing such data are complex to address those
problems, have large time complexity and consume a lot of memory.

Newer instruments still reduce sequencing costs and enhance
throughput and data quality, but they are still not ideal.
To reduce impact of the aforementioned difficulties
the experiments can be run with higher sequencing coverage (average number of reads containing a single base from a genome). 
It is also possible
to mix reads from various instruments to exploit different advantages of them,
e.g., short, but of good quality, Illumina reads can be used with long, intensely erroneous 
Pacific Biosciences reads. 
Alas, such approach often poses only a workaround of
the problem, since it causes another difficulties. Sequencing cost,
even though currently strongly decreased, is still considerable in limited scientific
budgets. The onefold cost of sequencing is enlarged by sustained costs of storing and sharing
huge amounts (even hundreds of GBs per single experiment) of a redundant data.\looseness=-1

The problem of sequencing errors can be partially solved
by involving specialized correction algorithms. 
In the recent years many of them were developed. The correctors detect potential errors and try to correct them, or sometimes
reject strongly damaged parts of the data by truncating or removing the reads.

Yang \emph{et al.} \cite{YCA13} classified the correction algorithms according to the approach of modeling the problem to: (\emph{i}) $k$-spectrum-based, (\emph{ii}) suffix-tree/array-based,
and (\emph{iii}) multiple-sequence-alignment-based.
The idea of the first category is to extract all fragments of reads of length~$k$ (\mbox{$k$-mers}) and make use of data redundancy as the majority of $k$-symbol-long fragments of the sequenced genome would be represented
by a number of $k$-mers in reads. 
The rare \mbox{$k$-mers} are deemed as erroneous and altered in the reads to the most similar,
but more frequent ones.
The suffix-tree/array-based algorithms also extract read substrings, but
stores them in suffix data structures, which allows to utilize different-length substrings simultaneously.
The multiple-sequence-alignment-based algorithms select from the input
files such reads, that seem to origin from the same genome fragments.
Then they perform multiple sequence alignment to match them each
other. It permits to find the consensus value of the particular read bases.\looseness=-1

In \cite{DD17} we proposed RECKONER, a $k$-spectrum-based error correction algorithm. 
We performed comparison of it and the state-of-the-art
algorithms: RACER \cite{IM13}, BLESS 2 \cite{HRHMC16}, Blue \cite{GDPB12},
Musket \cite{LSS13}, Lighter \cite{LFL14}, Trowel \cite{LMHHKW14},
Pollux \cite{MBM15}, BFC \cite{Li2015}, Ace \cite{IM13} from the group
of $k$-spectrum-based algorithms and Karect \cite{AKS15} from
multiple-sequence-alignment-based algorithms.

In this paper we present significantly faster and giving better corrections version of RECKONER.
Firstly, we reduced the time of processing of gzipped input files. Such files are typical in practice due to huge sizes of the input dataset.
Secondly, we improved the method of automatic $k$-mer length determination, which has positive influence on the quality of corrections if user will not define this parameter.
Thirdly, we redesigned the parallelization scheme to make better use of modern multicore processors.

The rest of the paper is organized as follows.
Section~\ref{sec:algorithm} presents the proposed algorithm.
In Section~\ref{sec:results} we show and discuss the results of the experiments.
The last section concludes the paper.

\section{Our algorithm}
\label{sec:algorithm}

\subsection{RECKONER workflow}
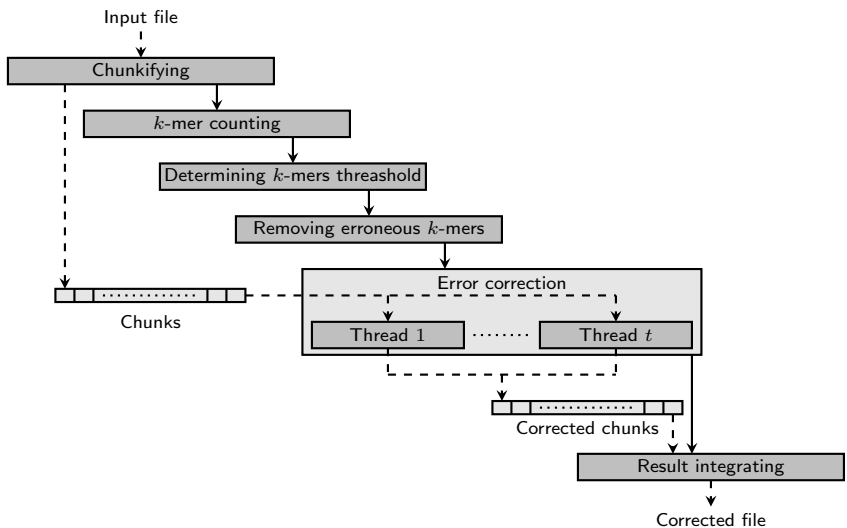
\begin{figure}[t]
\centering
\begin{tikzpicture}[>=stealth, x=0.5cm, y=0.35cm]

\newcommand{\Kolejka}[2]{
  \filldraw[black!10!white]($ (#1, #2) + (0, 0)   $) rectangle($ (#1, #2) + (5, 0.5) $);
  \draw[thick]         ($ (#1, #2) + (0, 0)       $) rectangle($ (#1, #2) + (5, 0.5) $);
  \draw[thick]         ($ (#1, #2) + (0.5, 0)     $)--($ (#1, #2) + (0.5, 0.5) $);
  \draw[thick]         ($ (#1, #2) + (1  , 0)     $)--($ (#1, #2) + (1, 0.5) $);
  \draw[dotted, thick] ($ (#1, #2) + (1.25 ,0.25) $)--($ (#1, #2) + (3.75, 0.25) $);
  \draw[thick]         ($ (#1, #2) + (4   ,0)     $)--($ (#1, #2) + (4, 0.5) $);
  \draw[thick]         ($ (#1, #2) + (4.5 ,0)     $)--($ (#1, #2) + (4.5, 0.5) $);
}

\newcommand{\Blok}[3]{
  \filldraw[black!25!white]($(#1,#2) + (0,0)$) rectangle($(#1,#2) + (7,1)$);
  \draw[thick]($(#1,#2) + (0,0)$) rectangle($(#1,#2) + (7,1)$);
  \draw($(#1,#2) + (3.5,0.5)$)[black] node{\sffamily\scriptsize #3};
}

\newcommand{\BlokMaly}[3]{
  \filldraw[black!25!white]($(#1,#2) + (0,0)$) rectangle($(#1,#2) + (4,1)$);
  \draw[thick]($(#1,#2) + (0,0)$) rectangle($(#1,#2) + (4,1)$);
  \draw($(#1,#2) + (2,0.5)$)[black] node{\sffamily\scriptsize #3};
}

\newcommand{\BlokDuzy}[4]{
  \filldraw[black!10!white]($(#1,#2) + (-0.25,-0.25)$) rectangle($(#1,#2) + (0,#4) + (10.25,1)$);
  \draw[thick]($(#1,#2) + (-0.25,-0.25)$) rectangle($(#1,#2) + (0,#4)  + (10.25,1)$);
  \draw($(#1,#2) + (0,#4) + (5,0.5)$)[black] node{\sffamily\scriptsize #3};
}


\draw(3.5,22.5)[black] node{\sffamily\scriptsize Input file};
\draw[dashed,->, thick] (3.5, 22) -- (3.5, 21);

\Blok{0}{20}{Chunkifying};
\draw[dashed,->, thick] (1.5, 20) -- (1.5, 12.25);
\draw[->, thick] (5.5, 20) -- (5.5, 19);

\Blok{2}{18}{$k$-mer counting};
\draw[->, thick] (7.5, 18) -- (7.5, 17);

\Blok{4}{16}{Determining $k$-mers threashold};
\draw[->, thick] (9.5, 16) -- (9.5, 15);

\Blok{6}{14}{Removing erroneous $k$-mers};
\draw[->, thick] (11.5, 14) -- (11.5, 13);

\Kolejka{1.25}{11.75};
\draw(3.75,11) node{\sffamily\scriptsize Chunks};

\BlokDuzy{8}{10}{Error correction}{2};
\BlokMaly{8}{10}{Thread $1$};
\BlokMaly{14}{10}{Thread $t$};
\draw[dotted,thick](12.25, 10.5) -- (13.75, 10.5);
\draw[->, thick] (18, 9.75) -- (18, 6);

\draw[dashed, thick] (6.25, 12) -- (16, 12);
\draw[dashed, ->, thick] (10, 12) -- (10, 11);
\draw[dashed, ->, thick] (16, 12) -- (16, 11);

\draw[dashed, thick] (10, 10) -- (10, 9);
\draw[dashed, thick] (16, 10) -- (16, 9);
\draw[dashed, thick] (10, 9) -- (16, 9);
\draw[dashed, ->, thick] (13, 9) -- (13, 8);

\Kolejka{12.75}{7.5};
\draw(15.25,7) node{\sffamily\scriptsize Corrected chunks};
\draw[dashed,->, thick] (17.5, 7.5) -- (17.5, 6);

\Blok{15}{5}{Result integrating}
\draw[dashed,->, thick] (18.5, 5) -- (18.5, 4);

\draw(18.5,3.5)[black] node{\sffamily\scriptsize Corrected file};

\end{tikzpicture}

\caption{RECKONER workflow; dashed lines denote data flow, solid lines denote data and control flow}
\label{fig:workflow}
\end{figure}

The workflow of original RECKONER is shown in Fig.~\ref{fig:workflow}.
Firstly, RECKONER traverses all reads present in the input files and
distributes them to groups (\emph{chunks}), that will be processed
separately by different threads. The positions of the first reads in chunks are
stored in a queue. Simultaneously, the minimum of read quality values is used for
determination of the level of quality coding. If it is lower than $59$ (the smallest value of
Phred+64 scale), it is supposed that the scale is Phred+33, and Phred+64 otherwise.\looseness=-1

Then KMC \cite{DKGD15}, is used to determine
the number of occurrences of all $k$-mers ($k$-mer \emph{counts}) and to build
a database of $k$-mers. The database is used to calculate a histogram of $k$-mer counts and
basing on this to determine the threshold of counts, which distinguish between probably erroneous and correct $k$-mers. The erroneous $k$-mers are
removed from the database by KMC tools~\cite{KDD17}.

The main correction is performed by many threads maintained by OpenMP. Every
thread picks a consecutive chunk from the queue. Then, it opens the input file
and performs correction of the reads of the current chunk. As a result it
saves to a temporary disk file description of introduced changes. For every
chunk one file is created. The threads work until any chunk is available.

In the last stage the results are integrated. The input file is traversed once
again. Now, according to the contents of consecutive temporary files, RECKONER
introduces changes to the reads, which are then stored in the output file.\looseness=-1

As we show in \cite{DD17}, RECKONER is one of the fastest read correction
algorithms. The experiments were, however, performed on uncompressed data. As
the stages of reads checking and results integrating are single-threaded, the time
of performing them constitute a significant fraction of the complete processing.
This effect is visible especially for compressed data, as in the last
stage the output reads have to be compressed by a single thread. Moreover,
the implemented method of chunks generation requires remembering of many positions in the input file.
For gzipped input the time of seeking these positions in a file could be
significant, especially when the files are large.
Due to huge size of the input
data, typically the only reasonable approach is to correct them without prior decompression, which enlarges the importance of addressing the problem. 

The following subsections present the improvements introduced to RECKONER to remove its drawbacks.

\subsection{Checking reads elimination}

To eliminate the necessity of one input file traversal we decided to
resign from the \emph{chunkifying} stage.
Instead, a short stage to determine the quality indicators level was added.
Currently it is made as follows.
A number of the first reads in the file
are taken until difference between one of quality
symbol value and the minimum value of Phred+33 ($33$) or the maximum
value of Phred+64 ($104$) is lower than $5$.
The level containing the found value is chosen.
Usually the number of reads necessary
to check is only a few.

To distribute the reads between threads we utilized a solution similar to the one used in KMC.
Before running the \emph{error correction} stage an additional thread is created.
It opens the input file and gets consecutive packages of data, fits them to contain
only the entire reads and places them in buffers. 
The correcting threads take the buffers and
correct the reads they contain.

\subsection{Result integrating optimization}

To optimize the last stage we decided to move the \emph{results integrating} into
the \emph{error correction} stage. The idea behind that was to perform the output compression
parallely. Currently, after performing correction of one chunk, the correcting
thread stores the corrected reads in a (usually compressed) temporary file.
The integration is performed in an additional thread,
which concatenates the previously generated files.
Correcting threads, after creating the temporary files, put information
about the ordering of the processed chunk to a priority queue, which guarantees the original
ordering of the chunks. 
Invariability of reads ordering in the output is required especially when processing paired-end reads.

\subsection{Other changes} \label{other_changes}
The original RECKONER has a simple algorithm for determination of $k$-mer length,
which is the most important parameter having a huge impact on quality of correction. 
The method is loosely based on logarithmic regression
with parameters chosen empirically. 
Unfortunately, the parameters were determined on experiments
with reads obtained with Mason \cite{H10}, which generates reads properly,
but their accompanying quality indicators are not adequate to the assumed quality
profile of the reads. As RECKONER utilizes the indicators intensively,
the results could be unreliable.
At this moment for read generation we use Art~\cite{HLMM12},
which accordingly to our observations generates the quality indicators properly.

Currently to compute the $k$-mer length we propose the following empirical
formula. It is based on results shown in Section~\ref{sec:results}:
$$
\begin{array}{cclcccl}
k&=&(a\log_2{g}-b)c, &\qquad\qquad &a&=&0.8,\\
b&=&9+p,					&				  &c&=&2+\ell/100,
\end{array}
$$
where $a$ and $b$ are logarithmic regression coefficients, $c$ is an adjustment
for different read length, $\ell$ is mean read length, $g$ is genome size
and $p$ is the average probability of error in \%; $g$ and $p$ are calculated with
the method shown in the supplementary material of \cite{DD17}. Moreover, $k$ is
lower- and upper-limited respectively by $k_{\min}=20$ and $k_{\max}=0.2\ell+30$.

\section{Experimental results}
\label{sec:results}
\subsection{Data sets and algorithms}
In \cite{DD17} we presented the tests of correctors efficacy for maize
genome, which is huge and highly repetitive. The conclusion was that
correctors work poorly for such data, some of them introduce even
more errors than properly correct. But as we said in Section
\ref{other_changes}, the employed read simulator (Mason) we used generated
unreliable quality indicators. Therefore in the present article we decided to reexamine the correctors
for maize reads obtained with Art~\cite{HLMM12}, which generates the indicators in compliance
with our expectations. Additionally we perform tests for human reads also obtained with Art.

We used our own profiles of
errors basing of the real read sets. Details of the profiles are given in
Table~\ref{tab:profiles}. For all datasets we set the sequencing coverage to~$20$. 
In \cite{DD17} we ranked the correctors in terms of both quality and resource requirements,
which ordered the algorithms starting from the best one as follows: RECKONER, BLESS, Blue,
Karect, Musket, BFC, Lighter, RACER, Ace, Trowel, Pollux. In this article we
selected RECKONER, BLESS, BFC, and Musket for comparison. We omitted Blue and Karect, as
they were not able to complete in  $12$ hours even for uncompressed data.

For RECKONER, BLESS, and Musket we selected $k$-mer length by performing preliminary experiments for a few values with step 3. The exact parameters and command
lines used for testing are shown in Appendix.
RECKONER is implemented in the C++11 programming language.
The experiments were run at workstation equipped with four AMD Opteron 6376 CPUs (16 cores each, clocked at 2.3\,GHz) and 512\,GB RAM running under openSUSE Leap 42.1 x86-64 OS.
The algorithms were run to use 64 threads. For compilation we used GCC 4.9.2.

\begin{table}[t]
  \caption{Data simulation profiles sources}
  \label{tab:profiles}
  \centering
  \renewcommand{\tabcolsep}{0.5em}
  \begin{tabular}{@{\extracolsep{0.45em}}cccc}
    \toprule
    Accession & Read   & Instrument & Average base \\
    number    & length & model      & error probability \\
    \midrule
    SRR1203044 & $100\text{bp}$ & HiSeq & 2.0\,\% \\
    SRR1802178 & $250\text{bp}$ & MiSeq & 0.3\,\% \\
    \bottomrule
  \end{tabular}
\end{table}

\subsection{Correction time}

\begin{figure}[t]
  \centering
  \begin{subfigure}{0.3\linewidth}
    \includegraphics[width=\linewidth]{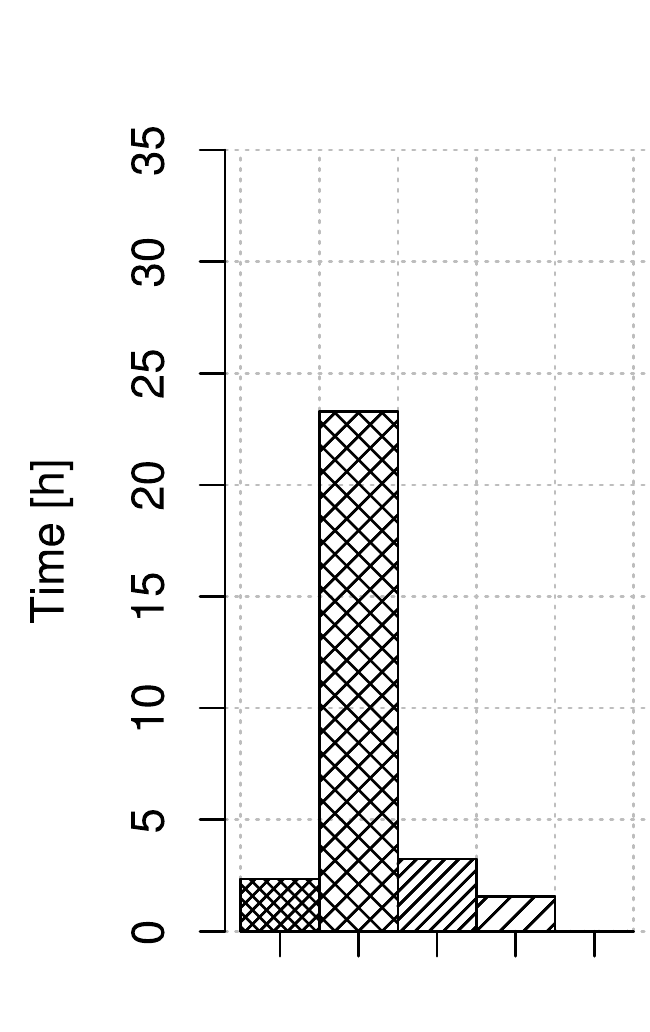}
    \caption{Maize}
    \label{fig:time:mays}
  \end{subfigure}
  \begin{subfigure}{0.3\linewidth}
    \includegraphics[width=\linewidth]{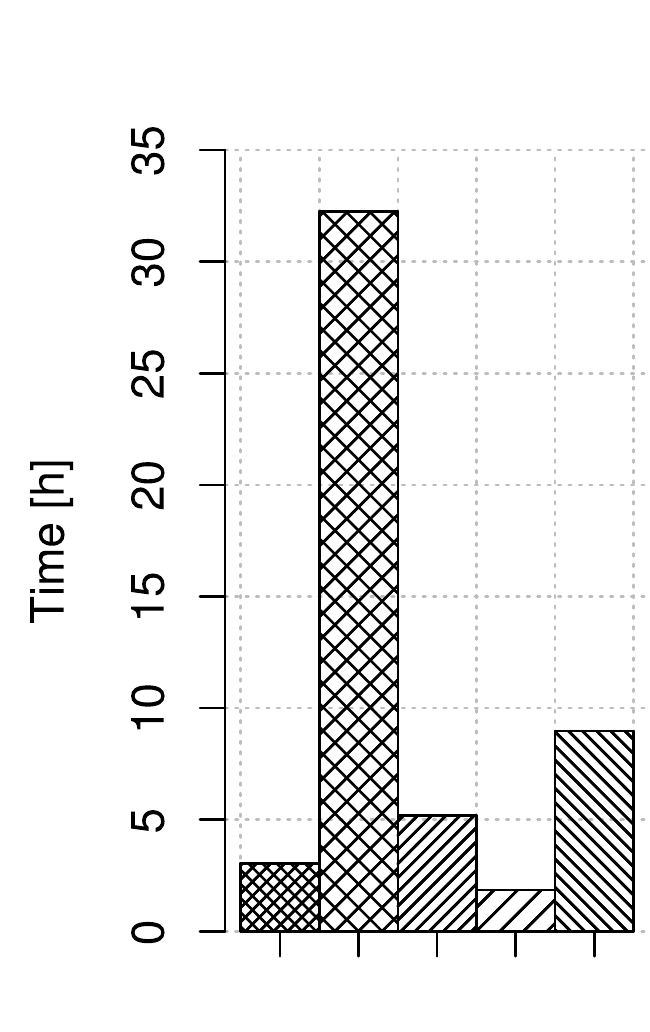}
    \caption{Human}
    \label{fig:time:sapiens}
  \end{subfigure}
  \begin{subfigure}{0.3\linewidth}
    \includegraphics[width=\linewidth]{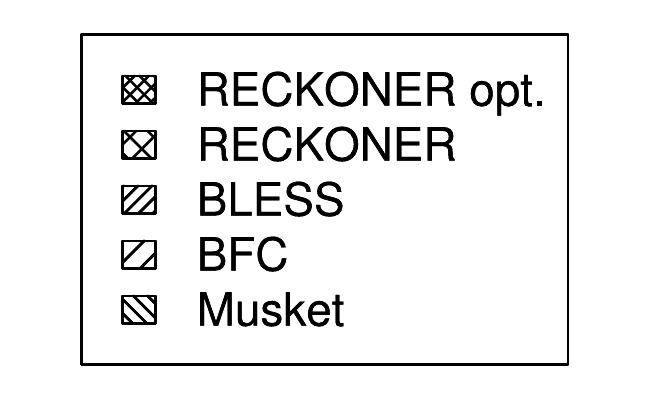}
  \end{subfigure}
  \caption{Correction time for Illumina HiSeq reads}
  \label{fig:time}
\end{figure}

We measured the time consumption of the correctors, including the optimized
version of RECKONER. Figure~\ref{fig:time} summarizes the results. For both
maize and human data, the time reduction for RECKONER is about 10-fold over its former version. Finally, the correction time reached 2--3 hours, which is definitely
acceptable for such huge organisms and is only a bit more than the
fastest algorithm, BFC. The missing bar of Musket in Fig.~\ref{fig:time:mays}
denotes that it was not able to correct the reads and returned untouched
input file.

What is important, the results were obtained for compressed input.
All correctors were able to read gzipped files, but only RECKONER and BLESS produced gzipped output files as well. BFC and Musket generated plain files. BLESS performs compression separately after the
correction finishes.
Thus, additional disk space (potentially huge) is necessary to store both uncompressed and compressed reads.

\subsection{Correction efficacy}

\begin{figure}[t]
  \centering
  \begin{subfigure}{0.3\linewidth}
    \includegraphics[width=\linewidth]{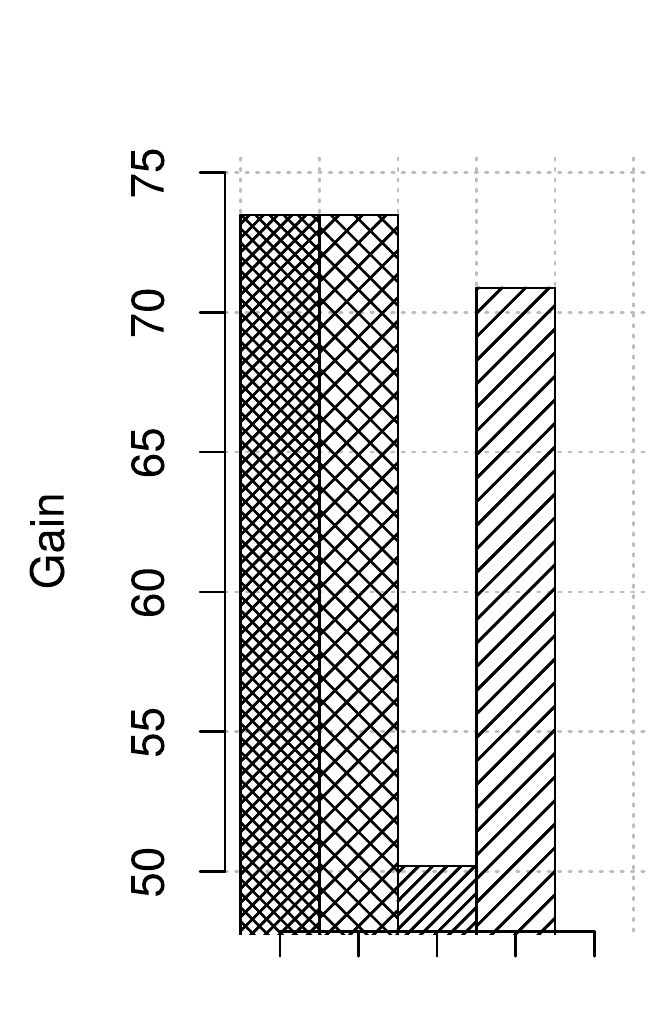}
    \caption{Maize}
    \label{fig:gain:mays}
  \end{subfigure}
  \begin{subfigure}{0.3\linewidth}
    \includegraphics[width=\linewidth]{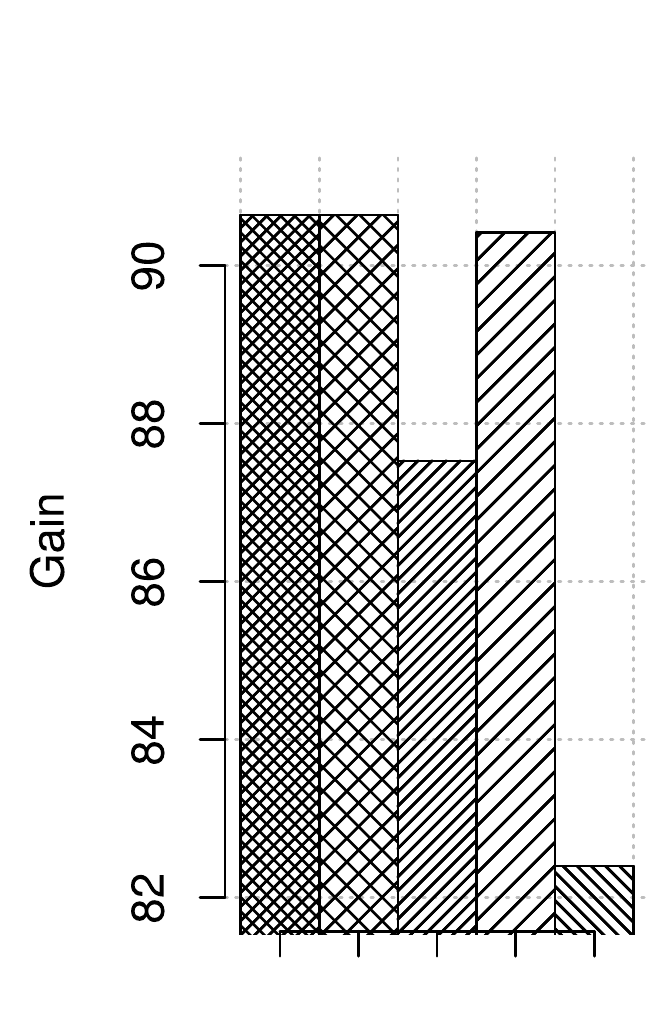}
    \caption{Human}
    \label{fig:gain:sapiens}
  \end{subfigure}
  \begin{subfigure}{0.3\linewidth}
    \includegraphics[width=\linewidth]{legend-eps-converted-to.pdf}
  \end{subfigure}
  \caption{Qualities of correction for HiSeq reads}
  \label{fig:gain}
\end{figure}

The corrector efficacy indicator, called \emph{gain} is defined as follows.
Let $\mathcal{TP}$ be a set of erroneous reads which were corrected perfectly,
$\mathcal{FP}$ be a set of error-free reads, which were disrupted by a corrector,
and $\mathcal{FN}$ be a set of erroneous reads, which were uncorrected at all or were miscorrected. Finally, 
$$\mathit{gain}={(|\mathcal{TP}|-|\mathcal{FP}|) / (|\mathcal{TP}|+|\mathcal{FN}|)}.$$

Figure~\ref{fig:gain} presents the results of quality in terms of gain. 
As no changes to the correction procedure are introduced, for the user-defined $k$-mer
length both old and optimized RECKONER obtain the same results. They are better
than for the competitors. 
The gains for maize data are worse than for human data, but they are approximately 2--3 times better than shown in~\cite{DD17}.

\subsection{MiSeq correction}

\begin{figure}[ht]
  \centering
  \begin{subfigure}{0.3\linewidth}
    \includegraphics[width=\linewidth]{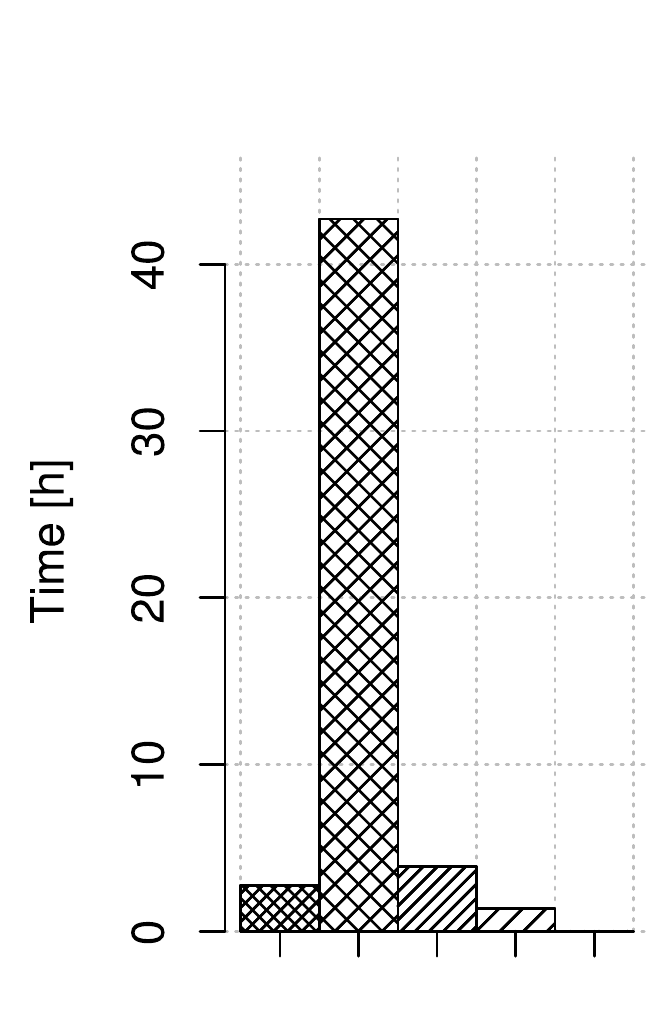}
    \caption{Time}
    \label{fig:miseq:time}
  \end{subfigure}
  \begin{subfigure}{0.3\linewidth}
    \includegraphics[width=\linewidth]{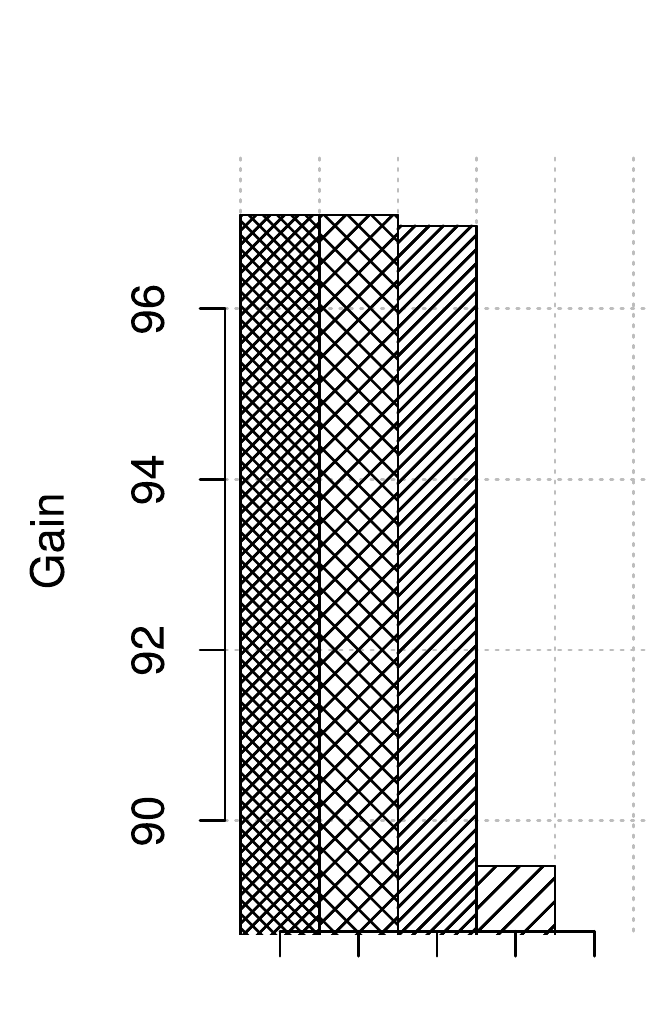}
    \caption{Gain}
    \label{fig:miseq:gain}
  \end{subfigure}
  \begin{subfigure}{0.3\linewidth}
    \includegraphics[width=\linewidth]{legend-eps-converted-to.pdf}
  \end{subfigure}
  \caption{Running times and qualities of corrections for MiSeq reads on human genome}
  \label{fig:miseq}
\end{figure}

In~\cite{DD17} we presented the results of correction only for HiSeq instruments.
As Illumina produces also other instruments, i.e., MiSeq, we performed tests for
it as well. Its important advantage are longer reads, in our case equal $250\text{bp}$. 
As MiSeq instruments produce low-error-rate reads, the profile with error rate 0.3\,\% was selected.
The computation times and quality results are shown in Fig.~\ref{fig:miseq}.

For these tests Musket also was unable to perform the correction.
For all other correctors the obtained gains are above 95 and very close. Thus, we can conclude that
the tested algorithms fit well also for longer reads. It is an important
asset, as MiSeq reads could be used together with HiSeq reads to improve
the quality of downstream algorithms by delivering longest reads.

\section{Conclusions}

In this paper we proposed new version of RECKONER, sequencing data correction algorithm.
The main improvements over the previous version are much faster processing and better selection of the key parameter, $k$-mer length.
The proper choice of $k$ is important for all $k$-spectrum-based correctors, as it has significant impact on the quality of correction.
The automatic determination of~$k$ is crucial in real-life applications as it is impossible to perform a number of corrections (for different values of $k$) to pick the best results.
The new version of RECKONER appears to be about ten times faster than its predecessor for gzipped input files which makes it one of the fastest algorithms.

We showed that $k$-spectrum-based algorithms are able to correct also
reads of highly-repeated genomes like maize. 
Moreover, we
showed, that such algorithms are able to correct also MiSeq reads.

\section*{Acknowledgement}	
The work was supported by the Polish National Science Center upon decision DEC-2015/17/B/ST6/01890.

\section*{Appendix}

To generate reads we placed a real source profile in a directory
\texttt{<profile\_dir>} and run read Art with the following commands:\\
\texttt{art\_profiler\_illumina <profile\_name> <profile\_dir>/ fastq}\\
\texttt{art\_illumina -sam -1 <profile\_name> -l <read\_length> -i <genome> -c <read\_number> -o <output> -rs 0 -na}\\

We run correctors with the following commands with the parameters specified
in Table~\ref{tab:correctors:params}:\\
\texttt{reckoner -kmerlength <k> -prefix . <input\_file>}\\
\texttt{bless -read <input\_file> -kmerlength <k> -prefix tmp -gzip}\\
\texttt{bfc -s <genome\_size> -t 64 <input\_file> > <output\_file>}\\
\texttt{musket -k <k> <kmers> -p 64 -o <output\_file> <input\_file>}

Sizes of the correctors input data is presented in Table~\ref{tab:sizes}.

\begin{table}[ht]
  \caption{Correctors versions and parameters}
  \label{tab:correctors:params}
  \centering
  \renewcommand{\tabcolsep}{0.3em}
  \begin{tabular}{cc|cccc}
    \hline\toprule
    \multirow{2}{*}{Algorithm} & \multirow{2}{*}{Version} & \multirow{2}{*}{Parameter} & Human & Maize & Human \\
     & & & HiSeq & HiSeq & MiSeq \\
    \midrule
    RECKONER & 1.0 & \texttt{k} & 36 & 42 & 78 \\
    \midrule
    RECKONER & 0.2.1 & \texttt{k} & 36 & 42 & 78 \\
    \midrule
    BLESS & 1.02 & \texttt{k} & 36 & 42 & 72 \\
    \midrule
    \multirow{2}{*}{BFC} & BFC-ht & \multirow{2}{*}{\texttt{genome\_length}} & \multirow{2}{*}{2991110000} & \multirow{2}{*}{2222330000} & \multirow{2}{*}{2991110000} \\
    & version v1 &&&&\\
    \midrule
    \multirow{2}{*}{Musket} & \multirow{2}{*}{1.1} & \texttt{k} & 33 & 33 & 72 \\
    & & \texttt{genome\_length} & 38649008822 & 29850675652 &  40692109119 \\
    \bottomrule
  \end{tabular}
\end{table}

\begin{table}[ht]
  \caption{Data sizes}
  \label{tab:sizes}
  \centering
  \renewcommand{\tabcolsep}{0.5em}
  \begin{tabular}{cccc}
    \hline\toprule
    Parameter & Human HiSeq & Maize HiSeq & Human MiSeq \\
    \midrule
    Gzipped file size & 50\,GiB & 38.9\,GiB & 42.9\,GiB \\
    Number of reads & 598,200,000 & 444,466,000 & 239,280,000 \\
    \bottomrule
  \end{tabular}
\end{table}

\end{document}